# Implementing a Web Browser with Phishing Detection Techniques

Aanchal Jain

Department of Computer Science And Engineering

Lakshmi Naraian College of Technology,

Bhopal(M.P.), India

aanchaljbpl@gmail.com

Prof. Vineet Richariya

Department of Computer Science And Engineering

Lakshmi Naraian College of Technology,

Bhopal(M.P.), India

vineet_rich@yahoo.com

*Abstract*—Phishing is the combination of social engineering and technical exploits designed to convince a victim to provide personal information, usually for the monetary gain of the attacker. Phishing has become the most popular practice among the criminals of the Web. Phishing attacks are becoming more frequent and sophisticated. The impact of phishing is drastic and significant since it can involve the risk of identity theft and financial losses. Phishing scams have become a problem for online banking and e-commerce users. In this paper we propose a novel approach to detect phishing attacks. We implemented a prototype web browser which can be used as an agent and processes each arriving email for phishing attacks. Using email data collected over a period time we demonstrate data that our approach is able to detect more phishing attacks than existing schemes.

Keywords- Phishing detection; Web browser.

## I. INTRODUCTION

The Internet is playing an increasingly significant role in today's commerce and business activities. Unfortunately, poor security on the Internet and large financial gains provide a strong motivation for attackers to perpetrate such seemingly low risk, yet high-return online scams. Email messages are not protected as they move across the Internet. Often information being transmitted is valuable and sensitive such that effective protection mechanisms are desirable in order to prevent information from being manipulated or to protect confidential information from being revealed by unauthorized parties.

Phishing can be classified into different types of attacks depending on the various channels of proliferation. These include malware, phishing emails, bogus Web sites, and identity theft. Malware is an application with malicious code that is distributed to the public via email or malicious Web sites. When victims access phishing emails or phishing Web sites, there is a chance that malware will be installed on the host computer and will steal personal information related to the customer surreptitiously [2]. Phishing email is another common type of phishing attack where phishers send out fraudulent emails impersonating genuine electronic service providers and ask victims to give away personal information or lead them to bogus Web sites. The bogus Web sites look similar to the Web sites of genuine electronic service providers. Once the victims log onto these Web sites, their personal information are recorded by the adversaries. Identity theft is the attack launched by phishers where they use the credentials of the victims to gain control of their accounts and transfer funds out of them.

Filtering phishing emails using classification techniques are able to control the problem in a variety of ways [14]. Detection and protection of phishing emails from the e-mail delivery system allows end-users to regain a useful means of communication. There are many different approaches for fighting phishing emails have been proposed. A promising approach is the use of content-based filters, capable of discerning phishing and legitimate email messages automatically. Many researches on content based email classification have been centered on the more sophisticated classifier-related issues. The success of machine learning techniques in text categorization has led researchers to explore learning algorithms in email classification [1]. Unlike most text categorization tasks, the cost of misclassification is heavily skewed. In order to address the growing problem, users and organizations analyze the tools with available to determine how best to counter phishing in its environment [5,6]. However, it is amazing that despite the increasing development of anti-phishing services and technologies, the number of phishing email messages continues to increase rapidly [5].

In this paper, we study the common practices involved in phishing attacks and review some anti-phishing solutions. We eventually focus on an approach which we have developed to detect and prevent phishing.

## II. BACKGROUND

Looking at the fact that phishing scammers are reaping enormous financial gains, it can easily be concluded that the motivation behind phishing is almost always financial. Although financial gain is the major motivating factor for phishing, other factors such as identity theft, industrial espionage, malware distribution, etc., are the other motivating factors for phishers. A root cause analysis was done to identify the motivation for phishing and the following factors were identified:

- Financial Gain
- Identity Theft
- Identity Trafficking
- Industrial Espionage
- Malware Distribution
- Password Harvesting
- Fame and Notoriety
- Exploit Security Holes

The causes for phishing vulnerability can be summarized into the following categories.

- Weak Authentication Schemes
- Browser vulnerabilities
- Security Flaws
- Non secure desktop tools
- Lack of user awareness
- Ease of impersonating a trusted source

There are currently several products available that use text classification to try to limit the potential damage caused by phishing. AntiPhish [4] is a browser extension which is used to protect inexperienced users against spoofed web site-based phishing attacks. AntiPhish is a plug-in tool which keeps track of users' sensitive information and prevents this information from being passed to a web site that is considered untrustworthy or unsafe. A text classification algorithm is responsible for identifying whether a web site is a phishing site based on addresses used in a form. It compares a legitimate URL and IP address with URL the page actually locates. AntiPhish focuses more on tracking sensitive information provided by a user. identified a website as a suspect phishing site when the visual similarity value is above a pre-defined threshold.

However, most phishing emails are sent asking the user to click on a hyperlink. After extensive analysis of hundreds of phishing emails and the methodologies used in phishing, phishing hyperlinks were categorized into the following general categories:

1) The actual link and the visual link in the email are different i.e., the hyperlink in the email does not point to the same location as the apparent hyperlink displayed to the users

2) The DNS name in the hyperlink is substituted by the quad-tuple IP address

3) DNS names used are manipulated to look similar to the genuine DNS name the phishers are trying to forge

4) The hyperlink is encoded so that it becomes very difficult to read for example, unusually long hyperlinks

5) When visiting the phishing hyperlink, it usually asks the user for various personal details like username, password, account number, SSN, etc.

The focus of this paper is to use the above mentioned data and use it to identify the possible phishing attacks. This approach will be integrated within the proposed web browser.

## III. METHOD

In our implementation, we experimented with link related feature:

1. visible_links: the total number of links in an emails.

2. invisible_links: total number of invisible links. This feature is calculated by an algorithm according to vision standard provided by W3C. In particular, if the color deference between the background and font of link in an email is less than 500, the link is considered as an invisible link.

3. Unmatching_urls: A binary value to show whether the visible url is as the same as the hidden url.

After defined the features that we want to look for in our algorithm, we developed a set of methods to extract all above mentioned three possible features from each email. The values of all features are numerical but in a different range. If we find the value for "invisible_links" and "unmatching_urls" to be nonzero then we consider the given email as a possible phishing attack. The proposed method can be illustrated in figure 1.

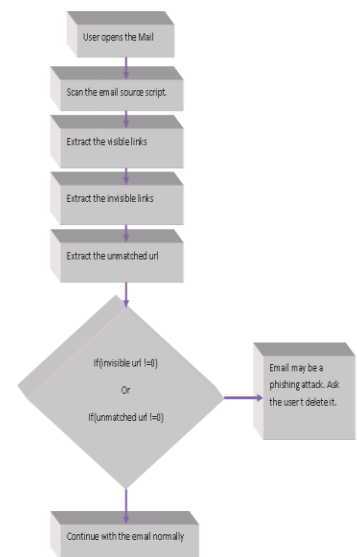

Figure 1: Flowchart of proposed method

## IV. IMPLEMENTATION

Our prototype currently includes a C#.Net implementation of a web browser. Installation of the system is simple. The user will have to use this web browser for opening the emails. A user will not see a difference for regular emails as the core of the browser is wrapped around Internet explorer's engine. In case an attack is detected the user is notified of the forged email suspicion and advised to delete the email. The proposed algorithm can be summarized as follows:

1. User opens the web browser and opens the email on the web browser.

2. The email before opening will be scanned by the backend phishing detection engine.

3. The "visible_links" will be extracted from the email body.

4. The "invisible_links" will be extracted from the email body.

5. The "Unmatching_urls" will be extracted from the email body.

6. If the count of "Unmatching_urls" or "invisible_links" is greater than 0.

   a. Prompt the user that this could be a a phishing attack.

   b. Advise him to delete the mail.

7. Else if

   a. Open the email normally.

   b. The status bar shows that the mail is marked as safe by our proposed phishing detection engine.

The prototype implementation of web browser displays the potential phishing attack as shown in the following screenshot:

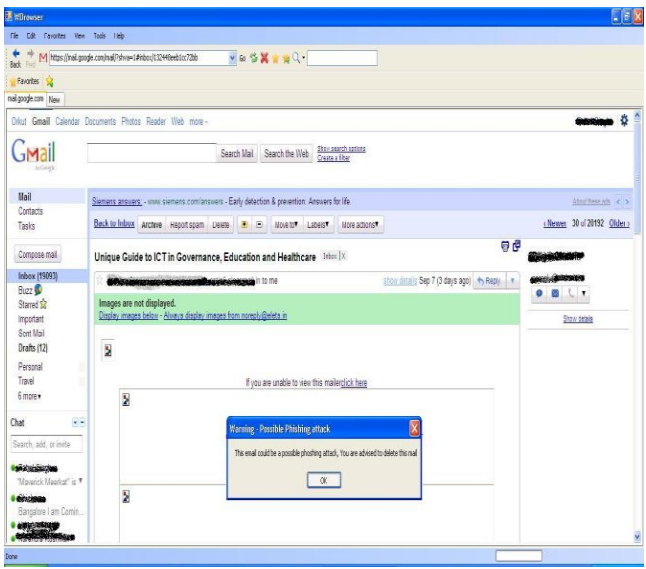

Figure 2: Screenshot of web browser with our method

## V. CONCLUSION

In this paper, we have presented an approach to detect phishing emails using link based features. The contribution of the work mainly consists of the usage of features visible links, invisible links and unmatched urls. The proposed algorithm used in conjunction with the proposed prototype of web browser will help the user to get notified of possible phishing attacks and will prevent them from opening the suspicious websites.

## ACKNOWLEDGMENT

We would like to thank all the supportive environment and staff of LNCT, Bhopal to give us the opportunity and facilities to carry out this work.

## AUTHORS PROFILE

**AANCHAL JAIN**

Completed Bachelor of engineering in Information Technology from RKDF Institute of science and technology , Rajiv Gandhi University, Bhopal (M.P.), India in 2009. Final Year Student Master of technology in Software Engineering (Dec'11) from Lakshmi Naraian college of Technology, Rajiv Gandhi University, Bhopal(M.P.), India

**Vineet Richariya**

Vineet Richariya is Head Of Department of Computer Science And Engineering, Lakhmi Narain College Of Technology, Bhopal, India. He did his MTech(Computer Science & Engineering) from BITS Pillali in year 2001.He did his B.E (Computer Science & Engineering) from Jiwaji University, Gwalior, India in year 1990.